\documentclass[10pt,journal,compsoc]{IEEEtran}
\usepackage{algpseudocode}

\usepackage{graphicx}
\graphicspath{ {pic/} }
\usepackage{float}
\usepackage{array}
\usepackage{tabularx}
\usepackage{balance}
\usepackage{marvosym}
\usepackage{amssymb}
\usepackage{amsmath}
\usepackage{multirow}
\usepackage{array}
\usepackage{enumitem}
\usepackage{booktabs} 
\usepackage[ruled,vlined]{algorithm2e}
\usepackage{amsmath,amssymb,amsfonts}
\usepackage{graphicx}
\usepackage{textcomp}
\usepackage{xcolor}
\usepackage{pifont}

\begin{document}
	
	\title{Enhancing Social Recommendation with \\Adversarial Graph Convolutional Networks}

	\author{Junliang~Yu, Hongzhi~Yin$^{*}$,
		Jundong~Li,
		Min~Gao,
		Zi Huang,
		and~Lizhen Cui
			\IEEEcompsocitemizethanks{\IEEEcompsocthanksitem J. Yu, H. Yin, and Z. Huang are with the School of Information Technology and Electrical Engineering, The University of Queensland, Brisbane, Queensland, Australia.\protect\\
			E-mail: jl.yu@uq.edu.au, h.yin1@uq.edu.au, huang@itee.uq.edu.au
			\IEEEcompsocthanksitem J. Li is with Department of Electrical and Computer Engineering, Department of Computer Science, and School of Data Science, University of Virginia, Charlottesville, Virginia, US.\protect\\
			E-mail: jundong@virginia.edu
			\IEEEcompsocthanksitem M. Gao is with the School of Big Data \& Software Engineering, Chongqing University, Chongqing, China.\protect\\
			E-mail: gaomin@cqu.edu.cn
			\IEEEcompsocthanksitem L. Cui is with the School of Software, Shandong University, Jinan, Shandong, China.\protect\\
			E-mail: clz@sdu.edu.cn
			}
		\thanks{$^{*}$Corresponding author and having equal contribution with the first author.}}

	\markboth{IEEE TRANSACTIONS ON KNOWLEDGE AND DATA ENGINEERING}%
	{Shell \MakeLowercase{\textit{et al.}}: Bare Demo of IEEEtran.cls for Computer Society Journals}

	\IEEEtitleabstractindextext{%
		\begin{abstract}
			Social recommender systems are expected to improve recommendation quality by incorporating social information when there is little user-item interaction data. However, recent reports from industry show that social recommender systems consistently fail in practice. According to the negative findings, the failure is attributed to: (1) A majority of users only have a very limited number of neighbors in social networks and can hardly benefit from social relations; (2) Social relations are noisy but they are indiscriminately used; (3) Social relations are assumed to be universally applicable to multiple scenarios while they are actually multi-faceted and show heterogeneous strengths in different scenarios. Most existing social recommendation models only consider the homophily in social networks and neglect these drawbacks. In this paper we propose a deep adversarial framework based on graph convolutional networks (GCN) to address these problems. Concretely, for (1) and (2), a GCN-based autoencoder is developed to augment the relation data by encoding high-order and complex connectivity patterns, and meanwhile is optimized subject to the constraint of reconstructing the social profile to guarantee the validity of the identified neighborhood. After obtaining enough purified social relations for each user, a GCN-based attentive social recommendation module is designed to address (3) by capturing the heterogeneous strengths of social relations. Finally, we adopt adversarial training to unify all the components by playing a Minimax game and ensure a coordinated effort to enhance recommendation performance. Extensive experiments on multiple open datasets demonstrate the superiority of our framework and the ablation study confirms the importance and effectiveness of each component.
		\end{abstract}
		
		}

	\maketitle

	\IEEEdisplaynontitleabstractindextext

	%
	\IEEEpeerreviewmaketitle

	\IEEEraisesectionheading{\section{Introduction}\label{sec:introduction}}
	
	\IEEEPARstart{R}{ecommender} systems are often compromised by the sparsity of data as users can only consume a tiny fraction of items, which leads to a frustrating result that in many cases recommender systems fail to fulfill their intrinsic capacity. With the popularity of social platforms, social relations are often used to in mitigate the problem of data sparsity since people's decisions are often influenced by their friends, according to the principle of \textit{homophily} \cite{Mcpherson2001Birds}. Suggested by this theory, a large number of social recommendation models \cite{Jamali2009TrustWalker,Ma2009Learning2,ma2011recommender,Guo2015TrustSVD,yin2016adapting,fan2019graph,wu2019neural,chen2020social}, which exploit explicit social information to improve general recommendation, have been developed, showing decent improvements under ideal circumstances. However, the real situations are rather complex and follow-up investigations \cite{Tang2013Social} from the industry revealed that the effectiveness of applying explicit social information into recommender systems had been overestimated.
	\par
	
	In summary, the failure of social recommendation can be attributed to: (1) The observed social relations are extremely sparse, and users with few ratings/purchase records are also likely to have few social relationships \cite{Zhang2017Collaborative}. Therefore it is questionable to claim that the majority of users who have limited purchase data can benefit from limited social relations. (2) The low cost of building social connections and the openness nature of social networks enable malicious accounts to easily infiltrate, making social networks noisy \cite{yu2017hybrid}. But social relations are often indiscriminately used in most social recommendation models.	(3) Social relations have multi-facets and show heterogeneous strengths in different situations \cite{tang2012mtrust} and are not universally applicable to any contexts. But most prior studies assume that users can reach a consensus with their friends in all the aspects of preferences. In view of these findings, it is less controversial that researchers should change their way of leveraging social relations and develop new approaches to better exploit the imperfect social network.
	\par
	Having been aware of the adverse factors of social recommendation, a few studies had made attempts to overcome these drawbacks. For example, to alleviate the influence of relation noises, a series of work \cite{Wang2016Social} aiming to distill the observable social networks was proposed in quick succession. Instead of using all the relations, they propose to extract useful explicit social links or divide social connections into fine-grained classes and only make use of the filtered information. However, in consideration of the high sparsity of original social networks, the extracted relations are too sparse to effectively improve recommendation performance. Meanwhile, an opposite way focusing on augmenting available relations was also explored. Researchers who follow this way paid attention to discovering potential but reliable relations for each user and identify those users who are not directly connected to the current user but share the similar interests with her as the \textit{implicit} friends \cite{Ma2013An,yu2018adaptive,hsu2018general}. 
	Following their ways, the sparsity of explicit social relations can be alleviated but on the other hand the augmentation also inevitably introduces noises. Specifically, in these methods, the friend-searching strategies are not coupled with the recommendation process, which questions the validity of the implicit friends for recommendation. Besides, none of these augmentation-based methods pay attention to the multi-facets problem of social relations	\par
	To overcome the limitations of prior research on social recommender systems. We propose to exploit the implicit social relations to enhance social recommendation. Technically, we present a deep adversarial framework (shown in Fig, 1) based on graph convolutional networks (GCN) \cite{zhou2018graph} to achieve our goals. We unfold our research by answering the following three questions:\\
	\textbf{(1) How does our framework address the problem of relation sparsity?} The studies focusing on the identification of implicit relations have set a good example for us that searching for unobserved but helpful connections upon the known social networks/bipartite networks is practicable \cite{Zhang2017Collaborative,yu2018adaptive,yu2019generating}. However, we claim that such models struggle to capture higher-order and complex connectivity patterns among nodes, because they only draw on information from immediate neighborhood or linear contexts (e.g. random walk based neighbors). Actually, in social networks, it has been shown to be beneficial to consider the higher-order structure \cite{benson2016higher}. In light of this, we follow their findings and further design some \textit{motifs} \cite{milo2002network} which are local structures involving multiple nodes to help uncover the high-order social information. To fully exploit the motifs, a tailored GCN is employed to aggregate information from motif-induced neighborhood and outputs a certain number of new relations, which are called the alternative neighborhood in this paper. By doing so, each user, especially those who have a very small number of relations, can benefit. \\
	\textbf{(2) How can our framework guarantee the validity of the alternative neighborhood?} The identification process of the new neighbors is unsupervised and hence it is hard to assess the validity of the alternative neighborhood before used for recommendation. Thus we adopt the idea of autoencoder and investigate if we can use these new neighbors to reconstruct the explicit social profiles. As shown in Fig. 1, we concatenate the motif-based GCN with a multilayer perception (MLP) and make them act as a concrete autoencoder \cite{abid2019concrete} for relation denoising. Technically, the GCN works as an encoder while the MLP acts as an decoder. By mapping the alternative neighborhood to part of the explicit social profiles, which are considered to be reliable, we believe that the informative and important information of the social network has been encoded into these the new neighbors. Intuitively, this design allows users to filter irrelevant relations or noises and guarantees the validity of the alternative neighborhood.
	\\
	\textbf{(3) How does our framework cope with the multi-facets problem of social relations?} After obtaining a certain number of helpful alternative neighbors for each user, we can replace the explicit relations with the new neighborhood. Then the multi-facets problem is the next to be handled. As the social relation may show heterogeneous strengths in different contexts \cite{tang2012mtrust}, it is necessary to develop a recommendation module that can selectively exploit the social information. To this end, we introduce an attentive social embedding propagation layer to our GCN based recommendation module in which a social attention mechanism is employed to weigh the contribution of the new neighbors and selectively aggregate information. Specifically, the attention mechanism is coupled with the context and then the importance of the social relations varies from context to context, enabling our framework to capture the heterogeneous strengths of trust to improve recommendation performance.
	\par
	In summary, our proposed framework includes three main stages: alternative neighborhood generation, neighborhood denoising, and attention-aware social recommendation, which are corresponding solutions to the above three questions. To unify all the tasks and components and ensure a coordinated effort against all the challenges, we finally adopt adversarial training \cite{goodfellow2014generative} to enhance our framework. In our adversarial framework, the motif-based GCN acts as a generator that produces the alternative neighborhood, while the attentive social recommendation module, works as the discriminator to recognize the informative connections and enforces the generator to learn the real distribution of the relations that can improve recommendation performance. With the competition between the generator and the discriminator, they mutually enhance each other. As a result, the whole framework will show better recommendation performance. Overall, the major contributions of this paper are summarized as follows:
	\begin{itemize}
		\item We design three different components to systematically and effectively deal with the three tough problems in social recommender systems. 
		\item We unify and intensify these three components by integrating them into a deep adversarial framework based on GCNs. To the best of our knowledge, we are the first to combine adversarial training and graph neural networks for social recommendation.	
		\item We conduct extensive experiments on multiple real-world datasets to demonstrate the superiority of the proposed framework and show the effectiveness of each component.
	\end{itemize} 
	The remainder of this paper is organized as follows. In Section 2, the related work is presented. Section 3 illustrates the proposed adversarial social recommendation framework. The experimental studies are presented in Section 4. Finally, Section 5 concludes the paper.

	\section{Related Work}
	\subsection{Social Recommendation} 
	The principle of homophily \cite{Mcpherson2001Birds} lays the theoretical basis for the majority of social recommendation models. Among the early attempts of social recommendation, the leading ideas can be categorized into three groups including co-factorization methods, ensemble methods, and regularization methods \cite{Tang2013Social}. As the representative work of co-factorization methods, SoRec \cite{ma2008sorec} and TrustMF \cite{yang2017social} jointly co-factorize the rating and social matrices and project rating and social contexts into the same latent space. STE \cite{Ma2009Learning} and mTrust \cite{tang2012mtrust}, as the typical ensemble methods, regard user's primary preference as the linear combination of ratings from the user and her social network. Another line of work, social regularization model \cite{ma2011recommender}, proposes to narrow the preference gap between users and their friends by using weighted social regularization terms. Besides, follow-up studies also explore social impact on users' preferences from other perspectives. The model of \cite{chen2018modeling} use social connections to capture users' exposure instead of preference. Yin \textit{et al.} \cite{yin2016adapting,yin2017spatial} integrated social relations into point-of-interest (POI) recommendation. SBPR \cite{zhao2014leveraging} leverages social connections to model the item ranks in personalized ranking. Group recommendation models \cite{yin2019social,yin2020overcoming} also borrow the strength of social relations to overcome the data sparsity in their scenarios.
	\par
	After the frustrating experience of applying social recommender systems in industry has been noticed \cite{Tang2013Social}, a few research efforts shift attention to addressing the practical problems of social recommendation. Based on the weak tie theory, Wang \textit{et al.} \cite{Wang2016Social} proposed fine-grained models to distinguish strong ties and weak ties. Inspired by the wide use of network embedding techniques, CUNE \cite{Zhang2017Collaborative} and IF-BPR \cite{yu2018adaptive} adopted homogeneous and heterogeneous network embeddings to capture reliable implicit social relations, respectively. InSRMF \cite{liu2019social} takes advantage of the indirect social relations detection and collaborative filtering on social network and rating behavior information to mitigate the relation sparsity. However, most of these studies only partially overcome the challenges of social recommendation.
	\subsection{Graph Neural Networks in Recommender Systems}
	Recently, graph neural networks (GNNs) \cite{zhou2018graph} have received increasing attention due to their great capacity in various tasks. Inspired by the success in other fields such as node classification and link prediction, researchers also investigated the suitability of GNNs for the task of recommendation. Representative work like NGCF \cite{wang2019neural} and LightGCN \cite{he2020lightgcn} have proven that GNN can significantly enhance general recommendation. As the message passing among nodes in GNNs is well aligned with the diffusion of social information, it is natural to adopt GNNs to social recommendation. Wu \textit{et al.} \cite{wu2019neural} developed a deep influence propagation model called DiffNet to simulate the recursive social diffusion process. Later, DiffNet++ \cite{wu2020diffnet++}, the extension of DiffNet, was proposed to model both the neural influence diffusion and interest diffusion in a unified framework. Besides, Fan \textit{et al.} proposed GraphRec \cite{fan2019graph} to learn user representations by fusing first-order social and first-order item neighbors. Wu \textit{et al.} \cite{wu2019dual}  developed a dual graph attention networks to collaboratively learn representations for two-fold social effects. Song \textit{et al.} proposed DGRec \cite{song2019session} to model both users’ session-based interests as well as dynamic social influences. Kim \textit{et al.} \cite{kim2019tripartite} proposed a graph-based social recommendation method that works on a group-user-item tripartite attributed multiplex heterogeneous graph to reduce noise and meanwhile paid attention to the oversmoothing problem of GNNs. Despite the effectiveness reported in these work, they either only address one of the problems discussed in Section 1, or neglect the limitations of directly using social information, resulting in a suboptimal performance.
	
	\begin{figure*}[t]
		\centering
		\includegraphics[width=0.93\textwidth]{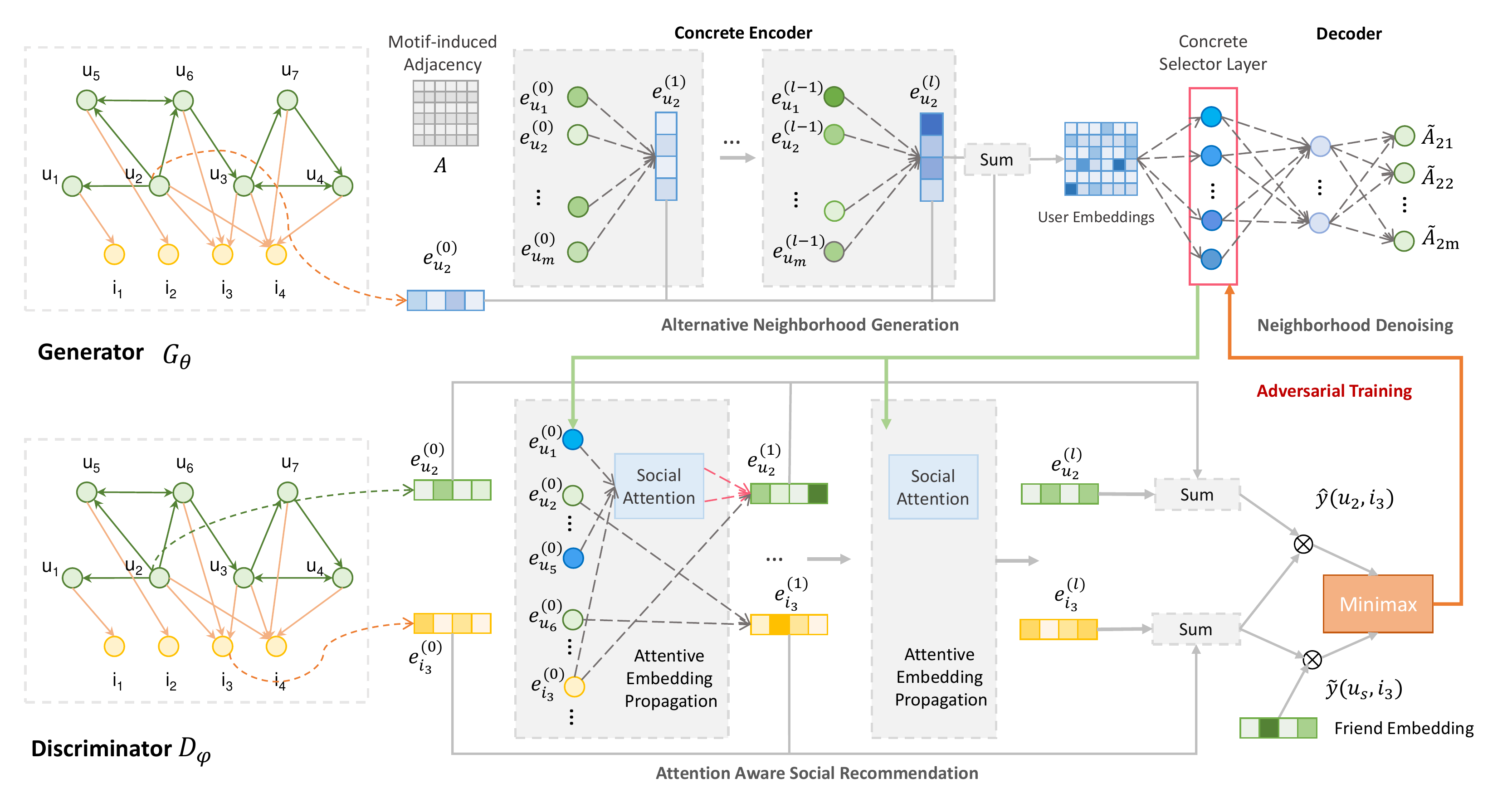}
		\caption{The schematic overview of our framework. The arrows with solid line indicate the flow of information and the arrows with dashed line represent the computational operations of the neural networks.}
		\label{figure.1}
	\end{figure*}

	\subsection{Adversarial Training in Recommender Systems} 
	Generative adversarial networks (GANs) \cite{goodfellow2014generative} have led a revolution in many fields including recommender systems. The most popular paradigm of applying adversarial training to recommender systems is that, by playing a Minimax game, the discriminator guides the generator towards fitting the underlying relevance distribution over items of the given user, while the generator generates difficult examples to confuse and then improve the discriminator \cite{wang2017irgan,wang2018graphgan,wang2018neural,chae2018cfgan,zhou2019adversarial}. IRGAN \cite{wang2017irgan} is the first GAN-based recommendation model which unifies the discriminative models and generative models. GraphGAN \cite{wang2018graphgan} designs a graph softmax function to accelerate training and improve computing efficiency. Wang \textit{et al.} \cite{wang2018neural} employed an adaptive negative sampling framework to optimize the proposed streaming recommendation model. Besides, instead of generating samples, Chae \textit{et al.} \cite{chae2018cfgan} proposed to learn user profiles with GAN. He \textit{et al.} \cite{he2018adversarial} made the first attempt to add adversarial perturbations to the latent factor to avoid overfitting. Additionally, \cite{wang2019minimax,wang2019enhancing} explore a new application of GAN by augmenting user-item interactions to improve collaborative filtering.\par
	There are also a few studies combining adversarial training and social recommendation \cite{krishnan2019modular,fan2019deep,yu2019generating}.  Krishnan \textit{et al} \cite{krishnan2019modular} proposed a modular adversarial framework to disentangle the architectural choices for the recommender and social representation models. Fan \textit{et al.} \cite{fan2019deep} adopt a bidirectional mapping method to transfer users' information between
	social domain and item domain using adversarial training. Among these work, RSGAN \cite{yu2019generating} is the most similar to our proposed model in terms of the motivation. This work also focuses on identifying reliable relations with adversarial training to improve social recommendation. However, it only covers the relation sparsity problem and the scarcity of labeled reliable friends for supervised learning is its bottleneck which limits the model's capacity.

	\section{Proposed Framework}
	In this paper we present our GCNs-based deep adversarial social recommendation framework. Figure 1 shows the schematic overview of our framework. We first present how each component in our framework deals with the aforementioned problems of social recommendation and then show how adversarial training unifies and intensifies all the components by playing a Minimax game. 
	\subsection{Notations and definitions}
	Let $U =\{u_1,u_2, ...,u_m \}$ and $I = \{i_1,i_2, ...,i_n\}$ respectively be the sets of users and items in recommender systems, where $m$ is the number of users, and $n$ is the number of items.  $\mathcal{N}(u)$ is the set of items which were clicked/consumed by user $u$ and $\mathcal{N}(i)$ is the user set including the users who clicked/purchased item $i$. $\mathbf{Y}\in \mathbb{R}^{m\times n}$ is the feedback matrix. For each pair $(u,i)$, $y_{ui}$ denotes the user's feedback on the item and $\hat{y}_{ui}$ denotes the predicted score. In this paper, we focus on Top-N recommendation. Following the convention, $y_{ui}$ is either 1 (positive) or 0 (negative or unknown). For each user pair $(u_{1},u_{2})$, $s_{u_{1},u_{2}}=1$ indicates that $u_{1}$ follows $u_{2}$ in the social network. It should be noted that we work on directed graphs and the relation matrix $\mathbf{S}\in \mathbb{R}^{m\times m}$ is asymmetric. The low dimensional vector $\mathbf{e}^{(l)}_{u_{1}}$ of dimension $d$ represents the node embedding where the superscript $l$ denotes the $l$-th layer and the subscript denotes the node. In this paper, we use bold capital letters to denote matrices and bold lowercase letters to denote vectors.
	\par

	\subsection{Alternative Neighborhood Generation}
	In this paper, alternative neighborhood refers to a group of users who share similar preferences with the specified user. Conceptually, it is literally like the \textit{implicit friends} proposed by \cite{Ma2013An, yu2018adaptive}, but actually a superset of them. For users with few social connections, the alternative neighborhood may be totally new neighbors, while partially overlaps with the neighbors of the users with a large number of social relations. We interchangeably call it alternative neighborhood/alternative neighbors/alternative relations in the remainder of this paper. Technically, prior studies \cite{Ma2013An,yu2018adaptive} identify implicit friends by means of rating similarity computation and network embedding techniques. But they struggle to capture the high-order and complex connectivity pattern among users because they only explore linear contexts (e.g. random walk based neighbors). In light of this, in our framework, we exploit the \textit{motifs} for alternative neighborhood identification. It should be noted that, the high-order information in existing GCN-based models such as NGCF \cite{wang2019neural} and DiffNet++ \cite{wu2020diffnet++} refers to the distant or multi-hop neighbors which are usually collected by Depth First Search. Being different from these existing literatures, the high-order in our work means relations beyond pairwise which cannot be captured with the conventional methods such as DFS.  \par
	\subsubsection{High-order Social Information Exploitation}
	Motif, which was first introduced in \cite{milo2002network}, is the specific local structure involving multiple nodes \cite{benson2016higher}, which are shown to be useful in many applications such as social computing \cite{milo2002network}. In this paper, we only focus on triangular motifs because of the widespread triadic closure in social networks. Fig. 2 shows the used triangular motifs in which $\mathcal{M}_{1} - \mathcal{M}_{7}$ are homogeneous and $\mathcal{M}_{8} - \mathcal{M}_{10}$ are heterogeneous. According to \cite{benson2016higher}, $\mathcal{M}_{1} - \mathcal{M}_{7}$ are crucial in social computing. To leverage the shared purchase preferences, we further design $\mathcal{M}_{8} - \mathcal{M}_{10}$ for our application. Specifically, the common purchases in $\mathcal{M}_{8}$ and $\mathcal{M}_9$ are used to strengthen the established relations, and $\mathcal{M}_{10}$ helps link users who are not directly connected via social relations. Obviously, introducing purchase information to social domain facilitates similarity measuring when social relations are very sparse.\par
	
	\begin{figure}[t]
		\centering
		\includegraphics[width=.45\textwidth]{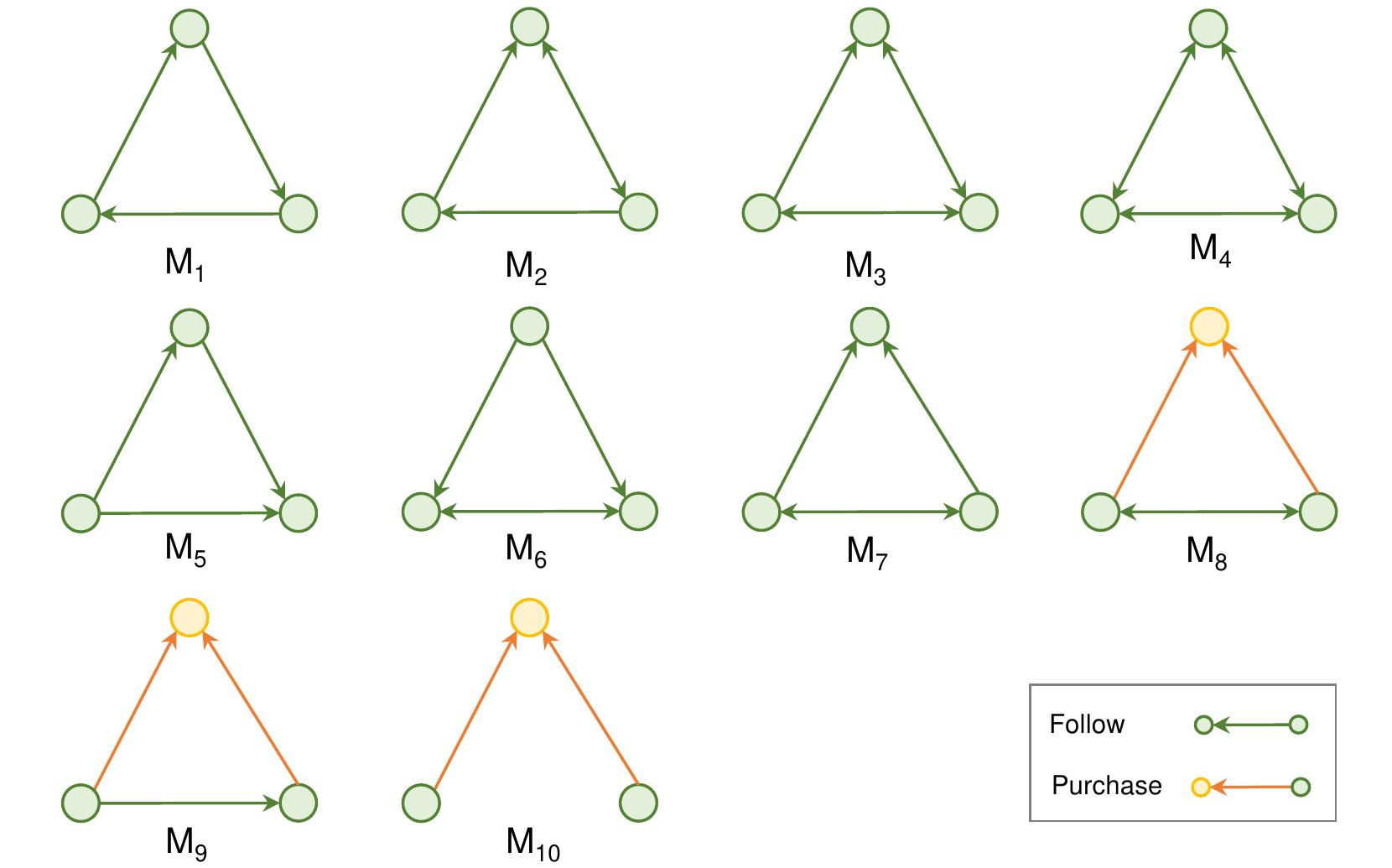}
		\caption{Motifs used in our work. The green circles denote users and the yellow circles denote items.}
		\label{figure.2}
	\end{figure}
	
	The motif-induced adjacency matrix $\mathbf{A}_{M_{k}}$ represents the frequency of two nodes appearing in a given motif $\mathcal{M}_{k}$, which is computed by:
	\begin{equation}
	(\mathbf{A}_{M_{k}})_{i,j}=\sum_{i\in U, j\in U} 1\ (i, j\ occur\ in\ \mathcal{M}_{k}).
	\end{equation}
	For example, given the motif $\mathcal{M}_{8}$, we can observe that, in the graph in Fig. 1, ($u_{3}, u_{4}, i_{4}$) forms one instance of  $\mathcal{M}_{8}$, thus $(\mathbf{A}_{M_{8}})_{u_{3},u_{4}}=1$. Counting motifs in a network may be time-consuming. But as we only exploit triangular motifs, following \cite{zhao2018ranking}, we show that motif-induced adjacency matrix can be computed based on simple matrix computation. Let $\mathbf{B}=\mathbf{S}\odot\mathbf{S}^{T}$ and $\mathbf{U}=\mathbf{S}-\mathbf{B}$ be the adjacency matrix of the bidirectional and unidirectional links of the explicit social network respectively, where $\odot$ denotes element-wise product. The computation of all the motif-induced adjacency matrices is summarized in Table 1.
	
	\begin{table}[!htb]
		\small
		\renewcommand\arraystretch{1.0}
		\caption{Computation of motif-induced adjacency matrices.}
		\label{Table:1}
		\begin{center}
			\begin{tabular}{c|l}
				\hline
				Motif&Matrix Computation  \\ \hline
				\hline
				$\mathcal{M}_{1}$ &$\mathbf{C}=(\mathbf{U} \mathbf{U}) \odot \mathbf{U}^{T}$ \\
				$\mathcal{M}_{2}$ & $\mathbf{C}=(\mathbf{B}  \mathbf{U}) \odot \mathbf{U}^{T}+(\mathbf{U} \mathbf{B}) \odot \mathbf{U}^{T}+(\mathbf{U}  \mathbf{U}) \odot \mathbf{B}$  \\
				$\mathcal{M}_{3}$&$\mathbf{C}=(\mathbf{B} \mathbf{B}) \odot \mathbf{U}+(\mathbf{B} \mathbf{U}) \odot \mathbf{B}+(\mathbf{U} \cdot \mathbf{B}) \odot \mathbf{B}$   \\
				$\mathcal{M}_{4}$ &$\mathbf{C}=(\mathbf{B} \mathbf{B}) \odot \mathbf{B}$ \\
				$\mathcal{M}_{5}$ &$\mathbf{C}=(\mathbf{U} \mathbf{U}) \odot \mathbf{U}+\left(\mathbf{U} \mathbf{U}^{T}\right) \odot \mathbf{U}+\left(\mathbf{U}^{T}  \mathbf{U}\right) \odot \mathbf{U}$  \\
				$\mathcal{M}_{6}$ &$\mathbf{C}=(\mathbf{U} \mathbf{B}) \odot \mathbf{U}+\left(\mathbf{B} \mathbf{U}^{T}\right) \odot \mathbf{U}^{T}+\left(\mathbf{U}^{T} \mathbf{U}\right) \odot \mathbf{B}$ \\
				$\mathcal{M}_{7}$ &$\mathbf{C}=\left(\mathbf{U}^{T} \mathbf{B}\right) \odot \mathbf{U}^{T}+(\mathbf{B} \mathbf{U}) \odot \mathbf{U}+\left(\mathbf{U}  \mathbf{U}^{T}\right) \odot \mathbf{B}$ \\
				$\mathcal{M}_{8}$ &$\mathbf{C}=\left(\mathbf{Y} \mathbf{Y}^{T}\right) \odot \mathbf{B}$ \\
				$\mathcal{M}_{9}$ &$\mathbf{C}=\left(\mathbf{Y} \mathbf{Y}^{T}\right) \odot \mathbf{U}$\\
				$\mathcal{M}_{10}$ &$\mathbf{C}=\left(\mathbf{Y} \mathbf{Y}^{T}\right) $\\
				\hline	
			\end{tabular}
		\end{center}
	\end{table}
	
	As can be seen, all the formulations are almost based on a general operation $(\mathbf{P} \mathbf{Q}) \odot \mathbf{T}$ (Here $\mathbf{P}$, $\mathbf{Q}$, and $\mathbf{T}$ are general notations denoting different matrices) which can be efficiently calculated since all the involved matrices are very sparse. For symmetric motifs, $\mathbf{A}_{M}=\mathbf{C}$ and for the asymmetric ones, $\mathbf{A}_{M}=\mathbf{C}+\mathbf{C}^{T}$. It is easy to understand these formulations as the operation $\mathbf{P} \mathbf{Q}$ constructs paths connecting the three vertices in a triangular motif and the operation $\odot \mathbf{T}$ complements the motif with the missing edge with a certain direction and closes the triangle. With the motif-induced adjacency matrices, we are able to leverage the high-order connectivity patterns in the social network and the user-item relations. Particularly, $\mathbf{A}_{M_{10}}$ is a complement that uses common purchase to establish implicit relations when users have few relations. It should be noted that we only preserve elements that are larger than 5 in $\mathbf{A}_{M_{10}}$ to avoid noises. \par
	
	\subsubsection{Neighborhood Generation with Motif-based GCN}
	Due to the great capacity to capture the dependence of nodes, we tailor the classical GCN \cite{zhou2018graph} to aggregate information from the explicit and motif-induced neighborhoods for the generation of the alternative neighborhood. Formally, a general GCN is constructed using the following layer-wise propagation \cite{zhou2018graph}:
	\begin{equation}
	\mathbf{E}^{(l+1)}=\sigma\left(\mathbf{Z} \mathbf{E}^{(l)} \mathbf{W}^{(l)}\right), \ \mathbf{Z}=\mathbf{D}^{-\frac{1}{2}} \tilde{\mathbf{A}} \mathbf{D}^{-\frac{1}{2}},
	\end{equation}
	where $\tilde{\mathbf{A}}=\mathbf{A}+\mathbf{I}$ ($\mathbf{I}$ represents an identity matrix with size $M$) with added self-loop, $\mathbf{D}$ is the diagonal degree matrix of $\tilde{\mathbf{A}}$, and $\sigma$ is the nonlinear activation function. However, it is reported in the latest studies \cite{he2020lightgcn,chen2020revisiting} that two most
	common designs in GCNs, feature transformation and nonlinear activation function, contribute little to the performance of collaborative filtering (CF). That is because GCN is originally proposed for node classification on attributed graph, where each node has rich attributes as input features, while in CF, each node (user or item) is only described by a one-hot ID and there is no concrete semantics behind the identifiers. In this paper, we following the design of LightGCN \cite{he2020lightgcn} and remove these two operations in the general GCN. As for the adjacency matrix, we defined it as follows:
	\begin{equation}
	\mathbf{A}=\mathbf{S}+\mathbf{A}_{M_{1}}+\cdots+\mathbf{A}_{M_{10}}.
	\end{equation}
	$\mathbf{S}$ is used because that motif-induced social network might cause some isolated nodes. To involve these isolated nodes which cannot build any motifs with other nodes, we need to incorporate the original relation adjacency matrix. At each layer, each user's node representation is refined by a weighted sum of the features of its motif-induced neighborhood. Formally, the neighborhood aggregation is defined as:
	\begin{equation}
	\begin{split}
	\mathbf{e}_{u_{i}}^{(l+1)}=\sum_{j\in\mathcal{N}_{(u_{i})}}a_{ij}\mathbf{e}_{u_{j}}^{(l)}.\\
	\mathbf{E}_{\theta}^{(l+1)}=\mathbf{D}_{A}^{-1}\mathbf{A} \mathbf{E}_{\theta}^{(l)}.
	\end{split}
	\end{equation}
	where $\mathbf{E}_{\theta}$ is the embeddings of all users. A more flexible solution may be allowing each node to choose the most beneficial motif-induced neighborhood to integrate information \cite{lee2019graph}. However, considering that our design has shown decent performance in the experiments, we leave the adaptive selection of motif-induced neighborhood as our future work.\par
	
	After propagating the high-order adjacency information through multiple layers, we obtain multiple representations of the the same dimension $d$ for each user, namely $\{\mathbf{e}_{u}^{(0)}, \mathbf{e}_{u}^{(1)}\cdots \mathbf{e}_{u}^{(l)}\}$. Each representation in this set captures different node interactions and semantics at different layers. To fully exploit the obtained user embeddings, we perform layer combination to get final representations $\mathbf{e}_{u}$ to predict the alternative neighborhood. Namely, $\mathbf{e}_{u}=\sum_{l=0}^{L} \frac{1}{L+1}\mathbf{e}_{u}^{(l)}$.  As our goal is to ensure the framework can be trained end-to-end, the process of neighborhood generation must be differentiable rather than using non-differentiable probability-based sampling. In other words, the output of the prediction should be discrete indexes, i.e., integers, that can represent user ID. To this end, a concrete selector layer \cite{abid2019concrete} is employed for discrete user selection. By using the concrete distribution \cite{maddison2016concrete} and the reparametrization trick \cite{kingma2013auto}, our framework is able to produce a relaxation of the one-hot vector to represent the selected new neighbor. The extent to which the one-hot vector is relaxed is controlled by a temperature parameter $\tau\in(0,\infty)$. In detail, the concrete selector layer contains $k$ neurons, each with a parameter $\mathbf{h}_{i}$ of size $m$, where the subscript $i$ denotes the index of the neuron. For each user, we conduct inner product between its final embedding and all users' final embeddings $(\mathbf{E_{\theta}}\mathbf{e}_{u_{i}}^{T})$ and then feed the intermediate result to the concrete selector layer. The layer performs such an operation $Softmax((\mathbf{E_{\theta}}\mathbf{e}_{u_{i}}^{T})\odot\mathbf{h}_{i})=\boldsymbol{\alpha}_{i}$ and then outputs a relaxation of one-hot vector, which is computed as:
	\begin{equation}
	\mathbf{v}_{i}=\frac{\exp \left(\left(\log \boldsymbol{\alpha}_{i}+\mathbf{g}\right) / \tau\right)}{\sum_{j=1}^{m} \exp \left(\left(\log \boldsymbol{\alpha}_{ij}+\mathbf{g}_{j}\right) / \tau\right)},
	\end{equation}
	where $\mathbf{g}$ is an $m$-dimensional vector of i.i.d. samples from $\mathit{Gumbel(0, 1)}$\footnote{$Gumbel(0,1)$ can be sampled with $g=-log(-log(\mu))$, where $\mu\sim Uniform(0,1)$.} \cite{gumbel1948statistical}, which is used to simulate the probability-based relation sampling. When $\tau\rightarrow0$, the concrete random variable $\mathbf{v}_{i}$ smoothly approaches the discrete distribution, and $v_{ij}$ would be 1 with the probability $\alpha_{ij}/\sum_{p=1}^{m}\alpha_{ip}$. The $j^{th}$ user would then be recognized as one of the new neighbors. As there are $k$ neurons in the concrete selector layer, we can total up all the output and then get a vector $\mathbf{v}$ with $k$ (e.g., 20) positions being 1 that denotes the alternative neighborhood. 
	\subsection{Neighborhood Denoising}
	Although a majority of users in social recommender systems have only a few friends, there are also a portion of users that have built a lot of connections with others. For these users, identifying alternative neighborhood is like relation denoising. The final identified alternative neighborhood should be those users which contribute most to the recommendation task. It is reasonable to think that, for users with a large number of social relations, if the alternative neighborhood can reconstruct the reliable part of their explicit social profiles, the essential pattern of connectivity may have been encoded into the alternative neighbors. In this paper, we use the motif $\mathcal{M}_{8}$-induced relations to serve as the reliable part because $\mathcal{M}_{8}$-induced relations are a subset of the explicit social relations while they are strengthened by common purchases. To impose this constraint, we concatenate the motif-based GCN with a fully-connected multi-layer perception (MLP) and make them nearly work as a concrete autoencoder \cite{abid2019concrete}. AutoEncoders have been widely used in feature selection to filter the irrelevant features \cite{li2017feature}. In our problem setting, filtering redundant and noisy relations for users with many social connections is equivalent to feature selection especially when we use the concrete autoencoder to enable discrete relation selection. Structurally, the GCN is the encoder while the MLP is the decoder. From the perspective of data flow, the input is the motif-induced adjacency matrix, the learned representation is the output of the concrete selector layer. Let $\mathbf{\tilde{A}}=f_{g}(\mathbf{A}_{M_{8}},\Theta)$ be the reconstructed motif-induced social profiles, where $\mathbf{A}_{M_{8}}$ is defined in Eq. 1 and $\Theta$ denotes the parameters of both the motif-based GCN and the MLP. We can formally define the objective function of the concrete autoencoder as:
	\begin{equation}
	\mathcal{L}_{s}=\arg\min \|\mathbf{A}_{M_{8}}-\mathbf{\tilde{A}}\|^{2}.
	\end{equation}
	By minimizing the above loss, the learned new neighborhood becomes less noisy even in the user preference domain due to the homophily across social and user-item networks. But it should be noted that, for the users who not only have few social relations but also few purchase records, the effect of the reconstruction constraint is limited as the supervision of a small number of ground truths is relatively trivial. In Section 3.5, we will continue discussing how to directly improve the validity of the alternative neighborhood in the user preference domain. 
	
	\subsection{Attentive Social Recommendation}
	With the alternative neighborhood, we next replace the explicit neighborhood for boosted performance. Due to the successful experience of GCN in practice \cite{fan2019graph, wang2019neural}, we are motivated to develop an attentive GCN-based recommendation model to perform social recommendation.
	\subsubsection{Attentive Social Embedding Propagation Layer}
	As social relations often show heterogeneous strengths in different situations, it is crucial to model the importance of social relations when propagating social embeddings. In our GCN-based recommendation model, we perform embedding propagation between both user pairs and user-item pairs. Given a user-item pair $(u,i)$, following \cite{he2020lightgcn}, we define the information passing from item $i$ to user $u$ at layer $l$ as:
	\begin{equation}\begin{aligned}
	\mathbf{e}_{u}^{(l+1)} &=\sum_{i \in \mathcal{N}_{u}} \frac{1}{\sqrt{\left|\mathcal{N}_{(u)}\right|} \sqrt{\left|\mathcal{N}_{(i)}\right|}} \mathbf{e}_{i}^{(l)} 
	\end{aligned}\end{equation}
	Analogously, the representation $e^{(l)}_{i}$ for item $i$ is obtained by aggregating information from users who purchased it. \par
	To capture the information from the alternative neighborhood when propagating, we propose a novel attentive social embedding propagation layer. Specifically, given a user pair $(u,v)$ where $v$ is one of the alternative neighbors, the user-to-user information aggregation is constructed as follows:
	\begin{equation}
	\mathbf{e}_{u}^{(l+1)}=\sum_{v\in \mathcal{A}_{u}}\alpha_{u,v}^{(l)}\mathbf{e}_{v}^{(l)}
	\end{equation}
	where $\mathcal{A}_{u}$ is user $u$'s alternative neighborhood. The difference between Eq (7) and Eq (8) is that we do not use the node centrality, e.g., degree, to measure the neighbor importance or how much information the alternative neighbor $v$ contributes. Instead, we use the learned weight $\alpha_{u,v}^{(l)}$ to selectively acquire information from the neighbor according to the historical records which act as the contexts when aggregating information. To calculate $\alpha_{u,v}^{(l)}$,  a social attention mechanism is employed and it formally operates in this way:
	\begin{equation}
	\alpha_{u,v}^{(l)}=\frac{\exp (\mathbf{q}^{(l)}\sigma(\mathbf{W}^{(l)}_{1}(\mathbf{e}_{u} + \mathbf{e}_{v}) || \mathbf{W}^{(l)}_{2}\mathbf{e}_{i}))}{\sum_{v'\in \mathcal{A}_{u}}\exp (\mathbf{q}^{(l)}\sigma(\mathbf{W}^{(l)}_{1}(\mathbf{e}_{u} + \mathbf{e}_{v'}) || \mathbf{W}^{(l)}_{2}\mathbf{e}_{i}))}	
	\end{equation}
	where $i$ is the item sampled from the historical purchase data of user $u$, $\mathbf{q}\in\mathbb{R}^{2d}, \mathbf{W}_{1}\in\mathbb{R}^{d\times d}$ and  $\mathbf{W}_{2}\in\mathbb{R}^{d\times d}$ are the parameters of the attention layer where $\mathbf{W}_{1}$ distills the user embedding and $ \mathbf{W}_{2}$ is for item embedding distillation, and $\sigma$ is the sigmoid function here. Although we are not the pioneer to apply graph attention mechanism to social recommendation, we are the first to couple attention with contexts to deal with the multi-facets problem of social relations. Compared with existing models, our attention mechanism has the advantage and novelty that the item embedding also participates in the computation of attention scores as the contextual information, which makes our attention mechanism better represent and capture the heterogeneity of social relations, while existing models compute attention scores just with the concatenation of two transformed user embeddings. Additionally, it should be noted that, as this work focuses on social recommendation, we only apply attentive embedding propagation to user-user connections for clear evaluation in the experimental part. \par
	Finally, we combine the social embedding and item embedding propagation to obtain the user representations of the next layer. The operation is defined as:
	\begin{equation}\mathbf{e}_{u}^{(l+1)}=\sum_{v\in \mathcal{A}_{u}}\alpha_{u,v}^{(l)}\mathbf{e}_{v}^{(l)}+\sum_{i \in \mathcal{N}_{u}} \frac{1}{\sqrt{\left|\mathcal{N}_{(u)}\right|} \sqrt{\left|\mathcal{N}_{(i)}\right|}} \mathbf{e}_{i}^{(l)}. 
	\end{equation}

	\subsubsection{Model Prediction and Optimization}
	By stacking $L$ attentive social embedding propagation layers, users and items are capable of receiving the information propagated from their $l$-hop neighbors. Following the design of LightGCN \cite{he2020lightgcn}, we combine different embeddings of different layers to constitute the final embeddings $\mathbf{e}^{*}_{u}$ and $\mathbf{e}^{*}_{i}$ for model prediction. Formally, $\mathbf{e}^{*}=\sum_{l=0}^{L} \frac{1}{L+1}\mathbf{e}^{(l)}$.  Given an instance $(u,i)$, the predicted score $\hat{y}(u,i)$ is computed by $\mathbf{e}^{*}_{u}\mathbf{e}^{*T}_{i}$. \par
	As this paper focuses on Top-N social recommendation, we adopt the pairwise ranking to model the order of items for each user. Following the Bayesian Personalized Ranking loss \cite{rendle2009bpr}, we define our optimization function as follows:
	\begin{equation}
	\mathcal{L}_{r}=\sum_{(u, i, j) \in \mathcal{O}}-\log \sigma\left(\hat{y}_{u,i}(\Phi)-\hat{y}_{u,j}(\Phi)\right)+\lambda\|\Phi\|_{2}^{2},
	\end{equation}
	where $\Phi$ denotes the parameters of the attentive recommendation module, and $\sigma(\cdot)$ here is the sigmoid function. Each time a triple including the current user $u$, the positive item $i$ purchased by $u$, and the negative item $j$ which is disliked by $u$ or unknown to $u$ sampled from the observed data $\mathcal{O}$, is fed to the model. The model is optimized towards ranking $i$ higher than $j$ in the recommendation list for $u$. In addition, $L_{2}$ regularization is imposed to reduce generalized errors.
	\subsection{Unifying All Modules with Adversarial Training}
	So far, we have built an end-to-end framework that can deal with the three tough problems in social recommender systems respectively. However, we notice that there is no distinct reward to stimulate the motif-based GCN to generate better alternative neighborhood that can further enhance recommendation performance. The neighborhood generation model and the recommendation model mainly focus on their own task rather than collaboration.  \par
	The theory of homophily refers to the tendency for people to have ties with others who are similar to themselves in socially significant ways. According to this theory, a good neighbor in social recommender systems is very likely to share the same preference with the current user. But for a given item only purchased by the current user, a reasonable assumption is that the neighbor would not show a higher level of interest on this item, which is analogous to the assumption of the social ranking model \cite{zhao2014leveraging} that a user tends to rank the items purchased by her higher than the ones purchased by her friends. Hence, we came up with such an idea that we should on one hand generate alternative neighbors who are interested in what the current user has purchased, but on the other hand make their affection to the items a little restrained. These two demands of our aim seem to be at cross-purposes. However, our goal is consistent with the key idea of the generative adversarial networks (GANs) \cite{goodfellow2014generative}. Inspired by the success of GANs in IR applications \cite{wang2017irgan,yu2019generating}, we are able to make the two demands of our aim compatible. \par
	The key idea of GANs is to let two neural networks contest with each other by playing a Minimax game. In our situation, we need to select new neighbors that can minimize the gap between the scores of the neighbors and the current user on the items purchased by the current user, which is explained to find neighbors who have the similar preferences with the current user. Meanwhile, we also have to maximize this gap because the user herself should show more affection to the purchased item. 
	Let $G$ denote the motif-based GCN, which plays the role of the generator, and $D$ denote the attentive recommendation module, which acts as the discriminator. The Minimax game under our framework can be formulated as:
	\begin{equation}
	\begin{split}
	\mathcal{L}_{adv}=\arg\min_{D_{\Phi}}\max_{G_{\Theta}}-\log\sigma(\hat{y}_{u,i}(\Phi)-\hat{y}_{u',i}(\Phi)), \\
	i\in\mathcal{N}_{u}, u'\in\mathcal{A}_{u}\sim P_{G}(\Theta|u, \mathbf{A}).
	\end{split}
	\end{equation}
	By fixing the parameters of $G$ and minimizing the above loss, $D$ is optimized towards recognizing the generated neighbor $u'$ and making the gap $(\hat{y}_{u,i}-\hat{y}_{u',i})$ larger. Oppositely, by fixing the parameters of $D$ and maximizing the loss, $G$ evolves towards generating neighbors that can narrow the gap. With the competition between $G$ and $D$, the optimization for $\mathcal{L}_{adv}$ will eventually reach an equilibrium where the framework shows the best performance. \par
	The use of adversarial training gives our framework a distinct objective that forces $G$ to generate new neighborhood which provides valuable information to $D$, and also empowers $D$ to capture fine-grained use preferences. Additionally, the neighbors who get lower score $\hat{y}_{u',i}$ would have less probability to be selected again, which can also be regarded as a way to denoise selected relations in the preference domain. With adversarial training, the interactions between the two GCNs in our framework increase, making all the components in our framework couple closely and ensuring a coordinated effort to improve social recommendation. Finally, we unify all the objective functions:
	\begin{equation}
	\begin{split}
	\mathcal{L}& = \mathcal{L}_{s}+\mathcal{L}_{r}+\mathcal{L}_{adv}\\
	& = \|\mathbf{A}_{M_{8}}-\mathbf{\tilde{A}}\|^{2}-\sum_{(u, i, j) \in \mathcal{O}}\log \sigma\left(\hat{y}_{u,i}(\Phi)-\hat{y}_{u,j}(\Phi)\right)\\
	& -\beta\arg\min_{D_{\Phi}}\max_{G_{\Theta}}\log\sigma(\hat{y}_{u,i}(\Phi)-\hat{y}_{u',i}(\Phi))\\
	\end{split}
	\end{equation}
	where $\beta$ is the coefficient to control the magnitude of the minimax game. For simplicity, we skip regularization terms here. The integrated objective function is optimized by Adam and all the parameters are jointly learned. As the training of GAN is highly unstable, we first pretrain the discriminator in a normal way (with $\mathcal{L}_{r}$) and pretrain the generator with $\mathcal{L}_{s}$ until the framework converges, then we conduct adversarial training to enhance the framework. Besides, it should be mentioned that, according to our experimental findings, strictly imposing $\mathcal{L}_{s}$ on the whole learning process may cause overfitting to explicit relations  and also result in mode collapse of the generator. Therefore, we just use it to pretrain the generator and fine tune the framework at the end of each iteration.
	\subsection{Complexity Analysis}
	\textbf{Model size}. As we follow the setting of LightGCN, we only need to learn the $0^{th}$ layer user and item embeddings of size $(m+n)d$ in graph convolution. Beside, for the MLP of the generator $G$, two matrices of size $m \times t$ are required for neighborhood denoising ($m$ is the number of users, and $t$ is the size of hidden units). The concrete select layer has the parameter of size $m\times k$. For the discriminator $D$, the weight matrices $\mathbf{W}_{1}$ and $\mathbf{W}_{2}$ for the attention mechanism are of the same size $d\times d$, and the parameter of $\mathbf{q}$ is of size $2d$. The total model size approximates $m(2t+k)+2Ld^{2}+(m+n)d$. Considering that $d, t$ and $k$ are small numbers generally less than 200 and the number of layers $L$ is usually not greater than 5, the model is fairly light. \par
	\textbf{Time complexity}. The computation cost is mainly from layer-wise propagation. For the two GCNs, without the attention mechanism, the layer propagation consumptions are $O(|\mathcal{A}^{+}|d)$ and $O(|\mathcal{Y}^{+}|d+kmd)$, respectively, where $|\mathcal{A}^{+}|$ is the number of non-zero elements in $\mathbf{A}$, $|\mathcal{Y}^{+}|$ is the number of non-zero elements in $\mathbf{Y}$, and $km$ denotes the number of total alternative neighbors. The time complexity for computation through fully connected layer in $G$ is $O(m^{2}t)$, but the denoising operation is not compulsory in each epoch. Involving the attention mechanism increases a certain amount of computation, but it also brings decent performance improvement. Hence, we need to make a trade-off between the better performance and time consumption in some cases to determine if the attention mechanism should be employed. As can be seen, in contrast to other GCN-based social recommendation models which directly use explicit social relations, our framework has extra time expenses in searching for alternative neighbors. However, considering the sparsity of the relation and feedback matrices, the compromise is acceptable. 
	
	\begin{table}[!htb]
		\small
		\renewcommand\arraystretch{1.1}
		\caption{Dataset Statistics}
		\label{Table:2}
		\begin{center}
			\begin{tabular}{c|cccc}
				\hline
				Dataset&\#Users & \#Items &  \#Feedbacks &  \#Relations   \\ \hline
				\hline
				LastFM &1,892 &  17,632 & 92,834 & 25,434  \\
				Douban & 2,848 & 39,586 & 894,887 &35,770 \\
				Gowalla&18,737  &32,510 & 1,278,274 &86,985\\
				\hline
			\end{tabular}
		\end{center}
	\end{table}
	
		\begin{table*}[t]
		\centering	
		\caption{General recommendation performance comparison.}
		\label{Table:3}
		\renewcommand\arraystretch{1.1}
		\begin{center}
			\begin{tabular}{cc|cccccccc|c}
				\hline
				Dataset&Metric&Random&BPR&SBPR&IF-BPR&DiffNet++&RSGAN&LightGCN&\textbf{ESRF}&Improv.\\ \hline
				\hline
				\multirow{3}{*}{LastFM}
				&Prec@10&8.113\%&7.585\%&8.276\%&9.102\%&10.083\%&9.475\%&10.139\%&\textbf{10.723}\%&5.848\%\\		
				&Recall@10&11.992\%&11.306\%&12.764\%&13.625\%&14.801\%&14.128\%&14.960\%&\textbf{16.076}\%&7.459\%\\			
				&NDCG@10&0.11826&0.10763&0.12188&0.13043&0.14316&0.13509&0.14371&\textbf{0.15341}&6.749\%\\
				
				\hline
				\multirow{3}{*}{Douban}
				&Prec@10&15.133\%&14.892\%&15.497\%&16.433\%&17.375\%&17.263\%&17.268\%&\textbf{18.231}\%&4.926\%\\		
				&Recall@10&5.225\%&5.023\%&5.422\%&5.542\%&6.136\%&6.032\%&6.095\%&\textbf{6.546}\%&6.681\%\\		
				&NDCG@10&0.16406&0.15855&0.17245&0.18551&0.19641&0.19469&0.19488&\textbf{0.21033}&7.087\%\\			
				
				\hline
				\multirow{3}{*}{Gowalla}
				&Prec@10&4.810\%&4.521\%&3.972\%&5.245\%&5.098\%&5.360\%&5.187\%&\textbf{5.842}\%&8.992\%\\			
				&Recall@10&5.755\%&5.484\%&4.807\%&6.287\%&5.993\%&6.327\%&6.102\%&\textbf{6.907}\%&6.906\%\\			
				&NDCG@10&0.06623&0.06311&0.05459&0.07279&0.07077&0.07450&0.07197&\textbf{0.08026}&7.731\%\\
				
				\hline
			\end{tabular}
		\end{center}
	\end{table*}
	
	\begin{table*}[t]
		\centering	
		\caption{Cold-start recommendation performance comparison.}
		\label{Table:4}
		\renewcommand\arraystretch{1.1}
		\begin{center}
			\begin{tabular}{cc|cccccccc|c}
				\hline
				Dataset&Metric&Random&BPR&SBPR&IF-BPR&DiffNet++&RSGAN&LightGCN&\textbf{ESRF}&Improv.\\ \hline
				\hline
				\multirow{3}{*}{LastFM}
				&Prec@10&5.906\%&5.903\%&6.129\%&6.857\%&6.787\%&7.142\%&6.806\%&\textbf{7.731}\%&8.246\%\\		
				&Recall@10&13.480\%&13.924\%&15.478\%&16.736\%&16.720\%&16.667\%&16.548\%&\textbf{18.410}\%&9.995\%\\			
				&NDCG@10&0.10674&0.11577&0.12128&0.13855&0.13901&0.14017&0.13127&\textbf{0.15230}&8.653\%\\
				
				\hline
				\multirow{3}{*}{Douban}
				&Prec@10&1.612\%&1.722\%&1.935\%&2.152\%&2.230\%&2.178\%&2.134\%&\textbf{2.421}\%&8.565\%\\		
				&Recall@10&6.735\%&7.178\%&8.084\%&8.441\%&8.705\%&8.583\%&8.317\%&\textbf{9.575}\%&9.994\%\\		
				&NDCG@10&0.04577&0.04784&0.05716&0.06246&0.06767&0.06447&0.06037&\textbf{0.07240}&6.989\%\\			
				
				\hline
				\multirow{3}{*}{Gowalla}
				&Prec@10&2.540\%&3.017\%&2.776\%&3.431\%&3.212\%&3.560\%&3.129\%&\textbf{3.993}\%&12.162\%\\			
				&Recall@10&5.528\%&6.693\%&5.855\%&7.285\%&6.854\%&7.427\%&6.995\%&\textbf{8.142}\%&9.627\%\\			
				&NDCG@10&0.04906&0.05782&0.05276&0.06262&0.05812&0.06487&0.05804&\textbf{0.07171}&9.538\%\\

				\hline
			\end{tabular}
		\end{center}
	\end{table*}
	\section{Experimental Results}
	\subsection{Experimental settings}
	\noindent\textbf{Datasets.} We evaluate our proposed model over three real-world datasets: LastFM\footnote{http://files.grouplens.org/datasets/hetrec2011/}, Douban\footnote{http://smiles.xjtu.edu.cn/Download/download\_Douban.html}, and Gowalla\footnote{https://www.dropbox.com/sh/qy3s8rs66nirhl9/AAClmTnFO-rR-4ecEYO-jU4ba?dl=0}. For the dataset of Douban, we only preserve the ratings larger than 3 as the user feedback because our aim is to generate Top-N recommendation. The details of the datasets are summarized in Table 2. Ten times of 5-fold cross validation is adopted in the experiments and we present the average results.\par
	\noindent\textbf{Baselines.} We name our framework \textbf{ESRF} (\textbf{E}hanced \textbf{S}ocial \textbf{R}ecommendation \textbf{F}ramework) in the experiments. To evaluate the performance of ESRF, we compare it with the following methods: BPR \cite{rendle2009bpr}, SBPR \cite{zhao2014leveraging}, IF-BPR \cite{yu2018adaptive}, RSGAN \cite{yu2019generating}, LightGCN \cite{he2020lightgcn} and DiffNet++ \cite{wu2020diffnet++}. Among them, BPR and LightGCN are general recommendation models based on implicit feedback while the rest are social recommendation models. As SBPR and DiffNet++ directly leverage explicit social relations for item ranking, comparing ESRF with them can validate the effectiveness of the alternative neighborhood.  IF-BPR and RSGAN are implicit relation-aware models which are highly relevant to ESRF, but they only partially overcome the intrinsic problems of social recommender systems. Choosing these two models as baselines can highlight the integrality of ESRF with regard to fully addressing the social recommendation issues. Specifically, LightGCN and DiffNet++ are GCN-based models as well.\par
	\noindent\textbf{Metrics.} To evaluate the performance of all methods, two relevancy-based metrics \textit{Precision@10} and \textit{Recall@10} and one ranking-based metric \textit{NDCG@10} are used. For a fair comparison, we rank all the candidate items to calculate these three metrics.\\
	\noindent\textbf{Settings.} In our experiments, we use grid search to tune all the specific parameters of baselines and ESRF to ensure the best performance of the baselines for fair comparison. For ESRF, the numbers of the alternative neighbors $k$ on LastFm, Douban, and Gowalla are set to 40, 30, and 40, respectively. The coefficients for controlling the magnitude of adversarial training $\beta$ are 0.3, 0.6, and 0.5, respectively. The temperature $\tau$ that controls the degree of relaxation for neighborhood generation is set to 0.2. For the general settings of all the models, we empirically set the dimension of latent factor vectors to 50, the regularization coefficient $\lambda$ to 0.005, and the batch size to 512. For all the GCN-based models, we choose the number of the layers according to their best performance on these datasets. The number of layers is set to 4, 3, and 4 on LastFM, Douban, and Gowalla, respectively. Section 4.5 further investigates the sensitivity of ESRF to the number of layers, and presents the results. For all the experiments, we use these settings.

	\begin{figure*}[h]
		\centering
		\includegraphics[width=.95\textwidth]{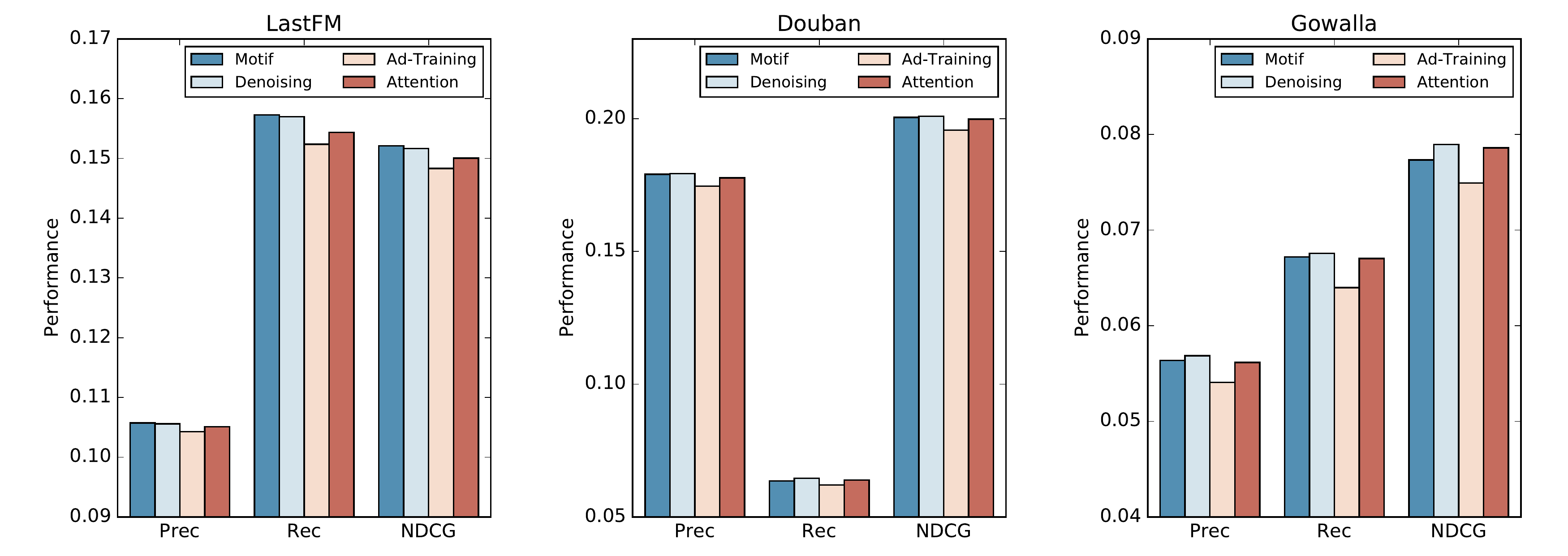}
		\caption{Ablation study on the three datasets.}
		\label{figure.3}
	\end{figure*}
	
	\begin{figure*}[t]
		\centering
		\includegraphics[width=.95\textwidth]{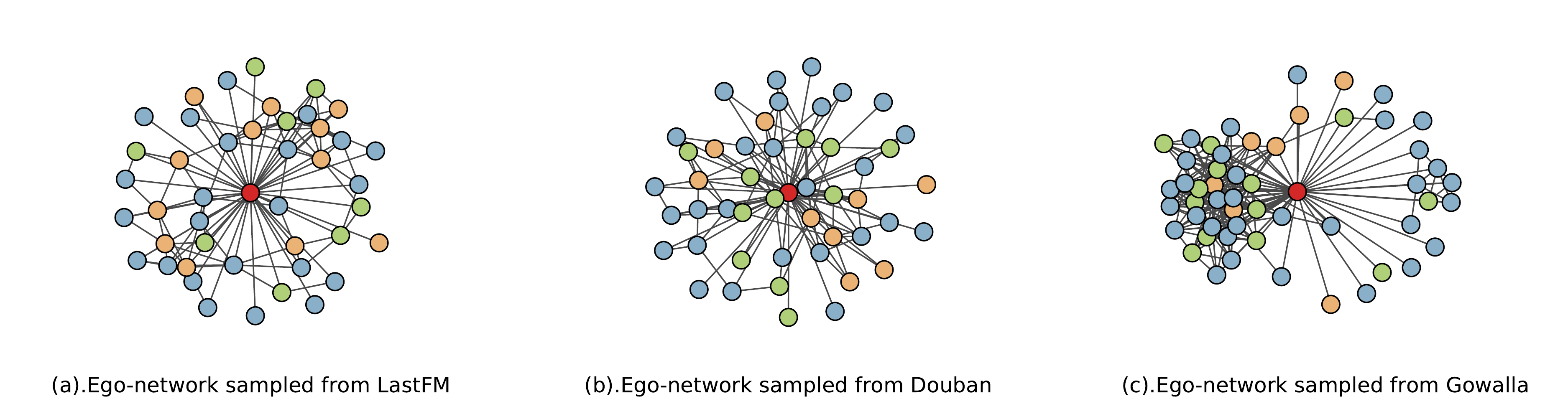}
		\caption{Three nodes with their ego social networks sampled from three datasets. The red node denotes the user herself, the blue nodes denote the explicit social relations, and the orange and green nodes denote the alternative neighbors. The overlapped part of explicit social relations and alternative neighbors is denoted by green nodes.}
		\label{figure.4}
	\end{figure*}
	\subsection{Recommendation Performance}
	In this part, we validate if ESRF is as effective as expected and outperforms the recent baselines. We respectively conducted experiments on the whole training set and the selected training set in which only the data of the cold-start users with less than 20 historical purchase records are included. The results are shown in Table 3 and Table 4.\par
	As can be observed, ESRF outperforms the baselines in both the general and cold-start cases on all the datasets. Particularly, the performance improvement in the cold-start case is remarkable, more than 9$\%$ on average over all the datasets. As for the baselines, we notice that GCN-based recommendation models: LightGCN and DiffNet++, outperform the corresponding shallow models: BPR and SBPR, which shows the great capacity of GCNs. Social recommendation models using explicit social relations by and large outperform (at least are comparable to) the general recommendation models on two denser datasets: LastFM and Douban. But when it comes to the dataset of Gowalla which is much sparser in both the social and preference domains, these social recommendation models are less competitive. This situation deteriorates in the cold-start case with SBPR even failing to beat BPR. But meanwhile, the implicit relation aware models: IF-BPR and RSGAN show good performance on three datasets, it is in line with the motivation of this paper that explicit social relations sometimes may mislead the social recommendation models and it is necessary to explore and exploit implicit but helpful relations. Besides, RSGAN which is also based on adversarial training shows the second or third best performance in many cases. The success of RSGAN and ESRF confirms the effectiveness of applying adversarial training to social recommendation. Finally, to verify that the the performance improvement of ESRF is brought by the generated alternative neighborhood, we randomly sample the same number of relations for each user and replace the alternative neighborhood with them in the discriminator of ESRF while suspend the generator. The modified ESRF is named Random in Table 3 and Table 4 and we can observe that the performance of Random is much worse than that of other models in all the cases. Then we can draw a conclusion that the alternative neighborhood contributes a lot to the performance of ESRF. 
	
	%
	%
	%

	\subsection{Ablation Study}
	In ESRF, we design three components to address three common problems in social recommender systems, and then use adversarial training to unify them. Here we respectively decouple the three components and adversarial training from ESRF to verify the effectiveness of each module. In Fig. 3, we present the results of modified ESRFs in different cases. In each case, one of the four modules of ESRF is removed from ESRF for general recommendation. Each bar represents a case that the corresponding module is removed. The lower the bar is, the more helpful the corresponding module is. It should be noted that we directly perform convolutional operation on explicit social relations to generated the alternative neighborhood when motifs are not used. As can be seen, each module contributes to the overall performance and the importance of modules varies from dataset to dataset. Overall, adversarial training is the most helpful part which accounts for the largest improvement compared with the baselines. This finding corroborates that our intuition of using adversarial generative networks is reasonable. The second important module is the social attention mechanism which plays a crucial role in graph convolution. Relatively speaking, the motifs and the denoising module contribute less but also beneficial. On the sparsest dataset - Gowalla, motifs show more importance than the denoising module, which can be attributed to its function to discover implicit connectivity pattern when the explicit connections are very limited. Besides, since adversarial training and the attention module can partially undertake the job of the denoising module, the missing of the denoising mechanism would not heavily degenerate ESRF. In summary, according to the results, we can draw a conclusion that capturing high-order social information is more essential than denoising when the social network is sparse, while adversarial training and social attention mechanism are always effective because they determine how the alternative neighborhood is generated and used.  
	
	\begin{figure}[t]
		\centering
		\includegraphics[width=.45\textwidth]{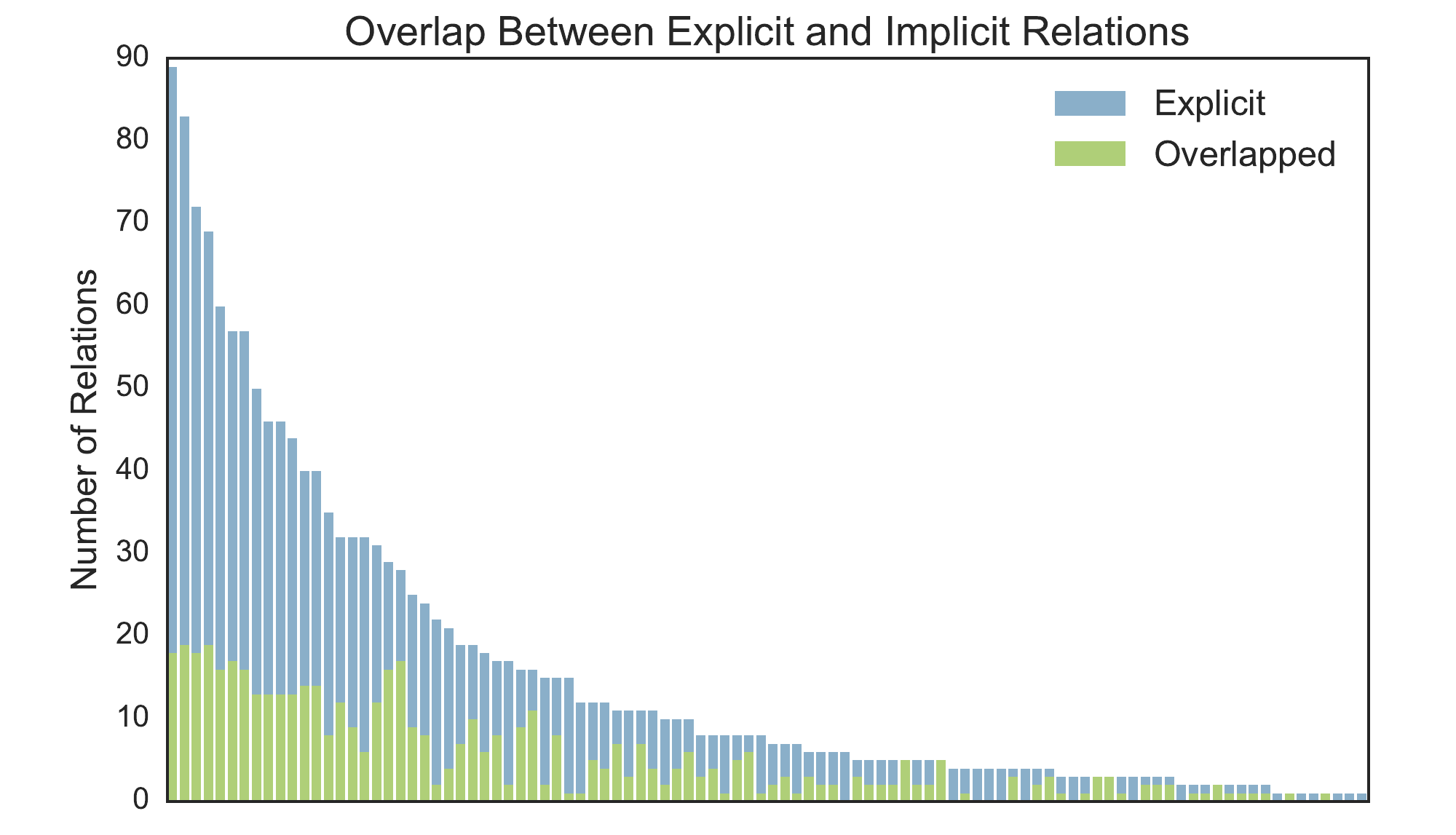}
		\caption{The degree of the overlap between explicit and implicit relations on LastFM.}
		\label{figure.5}
	\end{figure}
	\subsection{Comparison of Explicit Neighbors and Alternative Neighbors}
	As the focus of our paper, the alternative neighborhood is different from the explicit relations but has connections with them. To thoroughly investigate its characteristics, we first compare the connectivity patterns of explicit social neighbors and the alternative neighborhood. For each center user, we identify 20 alternative neighbors for her. Three ego social networks with both explicit and alternative neighbors are randomly sampled from three datasets, shown in Fig. 4. We can observe that: (1). More than half of alternative neighbors are also explicit neighbors when the center user has a large number of explicit social relations. (2). Compared with explicit neighbors, alternative neighbors are less isolated. Most of them are socially connected with other alternative neighbors or explicit neighbors. The second finding is very crucial. Intuitively, the strength of a tie becomes strong if the tie builders have common friends. According to the theory of homophily \cite{Mcpherson2001Birds}, they will also become similar in preferences. By contrast, the isolated explicit neighbors may have tenuous relationships with the center user and are less likely to share common interests with her. This can be a reason that explains why ESRF beats the social recommendation models using explicit social relations. Besides, we further investigate the degree of the overlap between the explicit relations and the alternative neighborhood. We randomly sample 100 users from LastFM and draw a distribution of the explicit relations. As can be seen in Fig. 5, more than 70\% users have less than 20 explicit relations and the distribution is a power-law-like distribution. For those users who have many explicit relations, we can observe an obvious overlap between the explicit relations and most of the identified alternative neighbors, which illustrates the alternative neighborhood's connections to explicit relations. Statistically, in this case, 32.44\% alternative relations overlap with the explicit relations. In addition, as we aim to capture the heterogeneous strengths of social relations, we propose a social attention mechanism to prevent relations being indiscriminately used. To validate its effectiveness, we collect the learned attention scores during training and randomly sample 20 users from LastFM and report the average results. In Fig. 6, each row in this heatmap represents a user and each column represents one of its identified 20 alternative neighbors. It is clear that, for each user, different alternative neighbors show different strengths in the convolution operation.

	\begin{figure}[t]
		\centering
		\includegraphics[width=.4\textwidth]{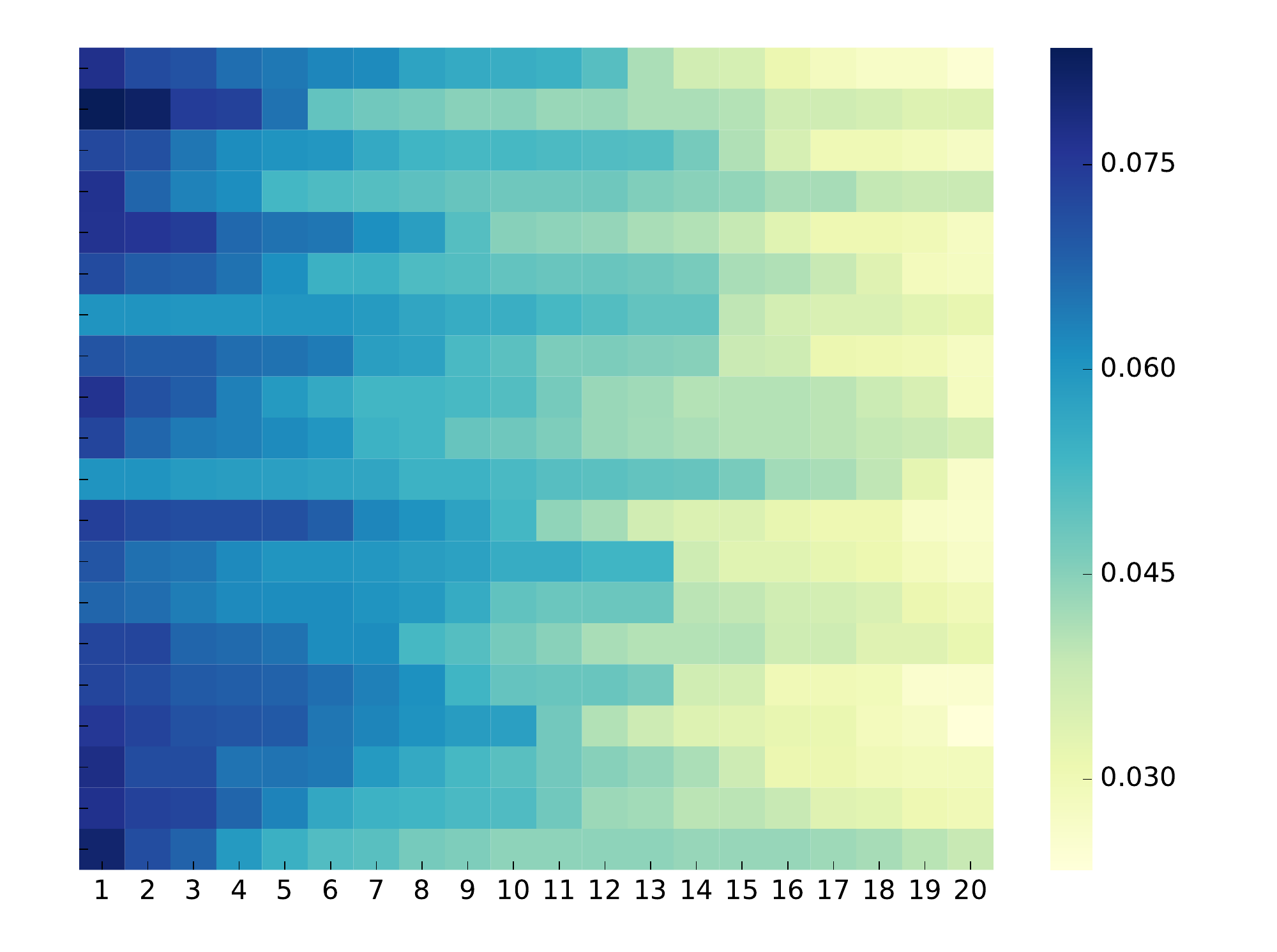}
		\caption{The heatmap of learned attention weights on LastFM.}
		\label{figure.6}
	\end{figure}

	\subsection{Parameter Sensitivity Analysis}
	We introduce two hyper-parameters $k$ and $\beta$ to ESRF. $k$ denotes the number of identified alternative neighbors for each user and $\beta$ controls the magnitude of adversarial training. In this section, we investigate the sensitivity of $k$ and $\beta$. As can be seen in Fig. 7, with the increase of the number of alternative neighbors, the performance of ESRF on all the datasets becomes better. After reaching its peak when $k$  is 40, 30, 50 on LastFM, Douban, and Gowalla, respectively, it steadily declines. Overall, the performance curve on Gowalla changes more dramatically, which could be attributed to Gowalla's high sparsity. These trends can be explained as that, augmenting the social information helps when data are sparse, but excessive augmentation may also introduce noises, which would mislead the model. From Fig. 8, we can observe similar trends that adversarial training improves ESRF, but a large magnitude of adversarial training will lower the performance of ESRF. Intuitively, we think it happens because maximizing the adversarial loss is opposite to our optimization goal. Excessive maximization would emphasize the goal of generating useful neighborhood while deprioritizing the recommendation task. Besides, as GCNs are sensitive to the number of layers, we also investigate the influence of the number of layers in this section. As the performance is mainly determined by the recommendation module and the effectiveness of motifs have been confirmed in Section 4.3, we only present the results on the discriminator in Fig. 9. As can be observed, with two layers, ESRF can almost outperform all the baselines (compare to Table 3/4). Increasing the number of layers can improve the performance, but the benefits diminish. The best performance is achieved when the layer number is 4, 3, and 4 in these three datasets, respectively.
	\begin{figure}[t]
		\centering
		\includegraphics[width=0.5\textwidth]{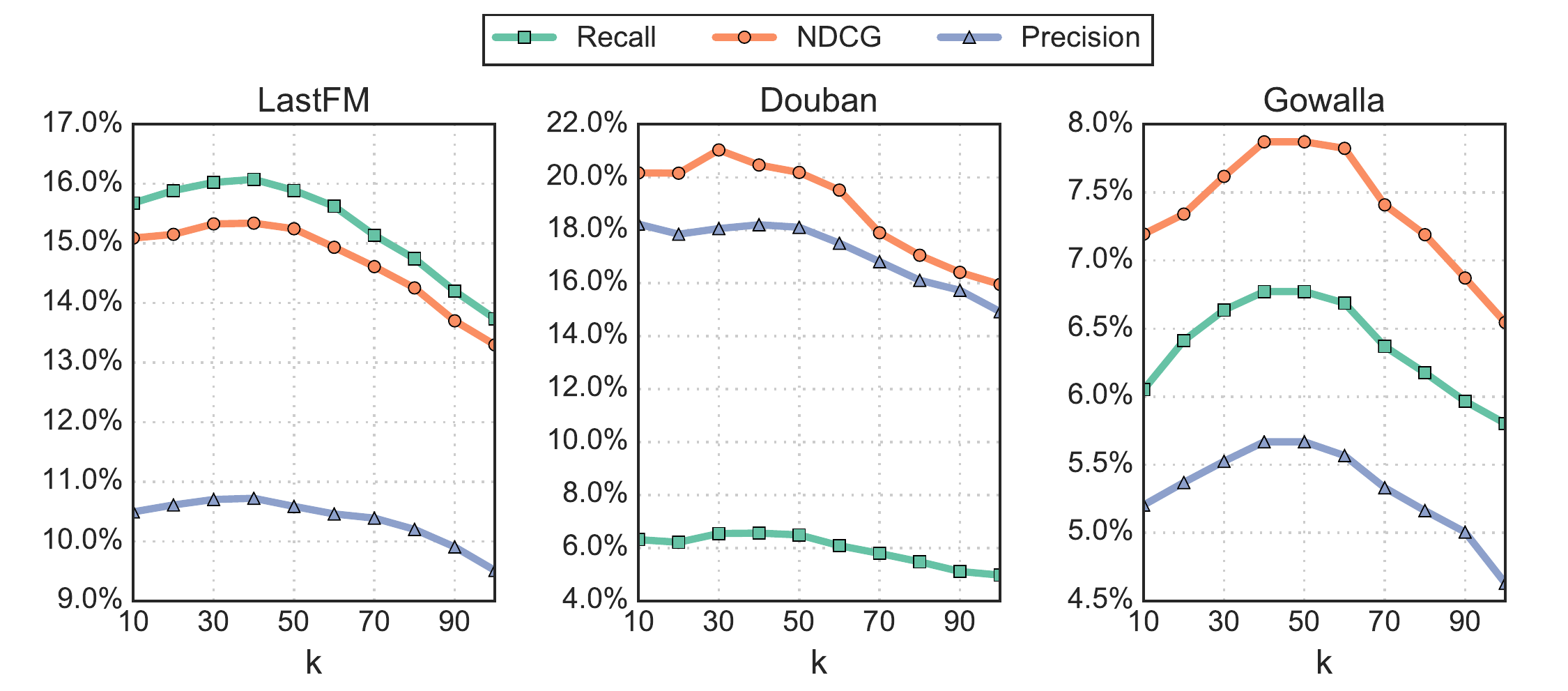}
		\caption{The influence of the number of alternative neighbors.}
		\label{figure.7}
	\end{figure}
	
	\begin{figure}[t]
		\centering
		\includegraphics[width=0.5\textwidth]{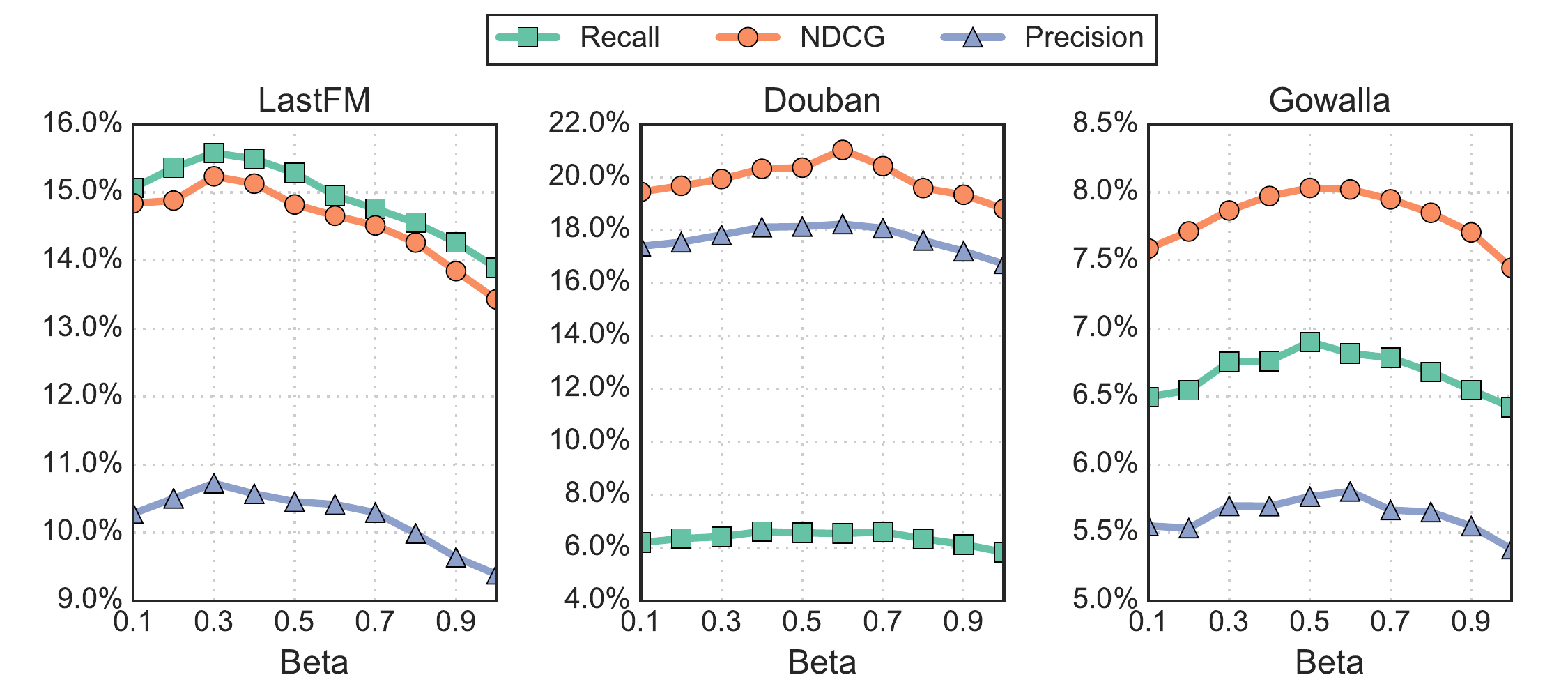}
		\caption{The influence of the magnitude of adversarial training.}
		\label{figure.8}
	\end{figure}
	
	\begin{figure}[ht]
		\centering
		\includegraphics[width=.5\textwidth]{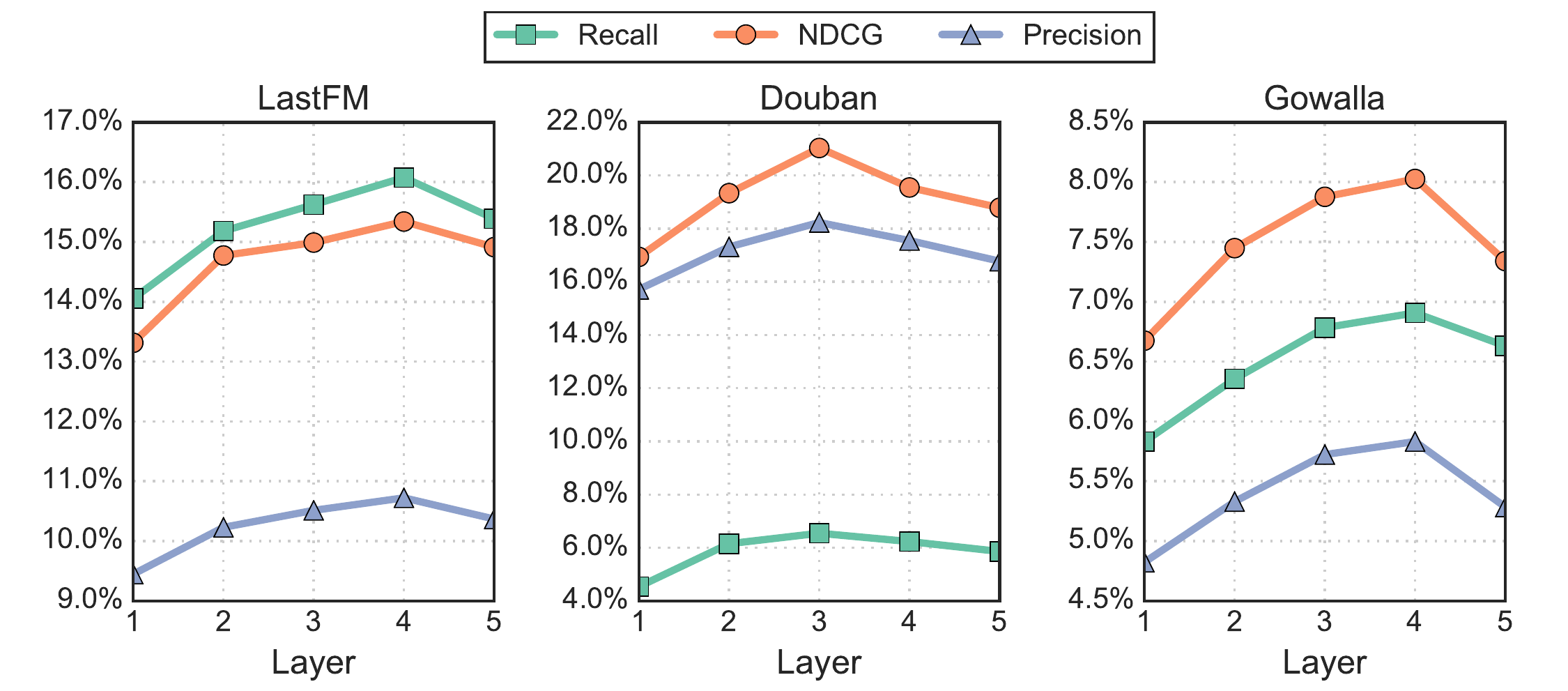}
		\caption{The influence of the number of layers of the discriminator.}
		\label{figure.9}
	\end{figure}
	
	\section{Conclusion}
	Social recommender systems have received attention these years due to the potential of social relations to improve traditional recommendation. However, the practice of social recommendation is not as successful as expected. Existing social recommender systems only pay attention to the homophily in social networks and neglect the intrinsic problems of social relations. In this paper, we develop a deep adversarial framework based on GCNs to address the challenges of social recommendation. To the best of our knowledge, this is the first work that combines adversarial training and GCNs to overcome the drawbacks of social recommender systems. Experimental results on three real-world datasets show that our recommendation model significantly outperforms other state-of-the-art social recommendation models. 
	
	\section{Acknowledgement}
	This work was supported by ARC Discovery Project (Grant No.DP190101985, DP170103954).

	\ifCLASSOPTIONcaptionsoff
	\newpage
	\fi
	
	\bibliographystyle{IEEEtran}
	\bibliography{refs}
	
	%
	\begin{IEEEbiography}[{\includegraphics[width=1in,height=1.25in,clip,keepaspectratio]{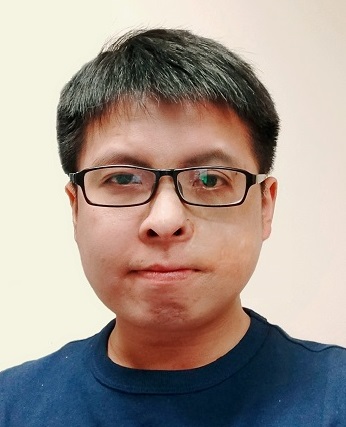}}]{Junliang Yu}
		received the B.S. and M.S degrees in Software Engineering from Chongqing University, Chongqing, China. Currently, he is a Ph.D. student with the school of Information Technology and Electrical Engineering at the University of Queensland, Queensland, Australia. His research interests include recommender systems and anomaly detection.
	\end{IEEEbiography}
	
	\begin{IEEEbiography}[{\includegraphics[width=1in,height=1.25in,clip,keepaspectratio]{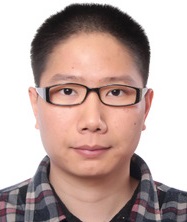}}]{Hongzhi Yin}
		received the PhD degree in computer science from Peking University, in 2014. He is a senior lecturer with the University of Queensland. He received the Australia Research Council Discovery Early-Career Researcher Award, in 2016 and UQ Foundation Research Excellence Award, in 2019. His research interests include recommendation system, user profiling, topic models, deep learning, social media mining, and location-based services.
	\end{IEEEbiography}
	
	\begin{IEEEbiography}[{\includegraphics[width=1in,height=1.25in,clip,keepaspectratio]{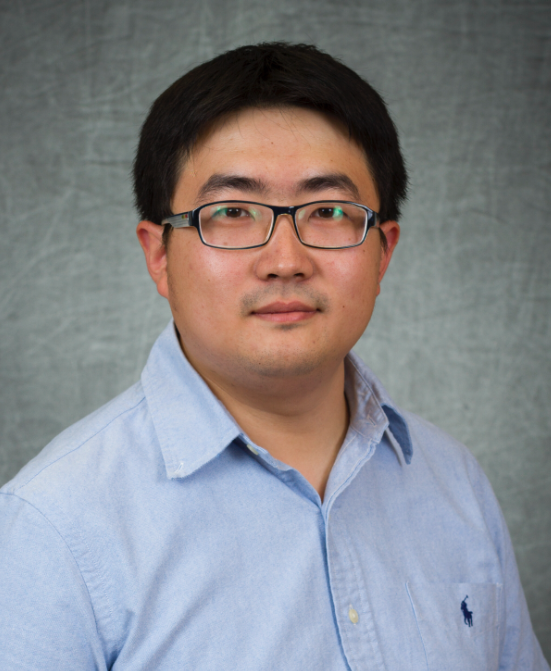}}]{Jundong Li}
		is an Assistant Professor at the University of Virginia. Prior to joining UVA, he received his Ph.D. degree in Computer Science at Arizona State University in 2019, M.Sc. degree in Computer Science at University of Alberta in 2014, and B.Eng. degree in Software Engineering at Zhejiang University in 2012. His research interests are broadly in data mining and machine learning, with a particular focus on feature learning, graph mining, and social computing.
	\end{IEEEbiography}
	
	\begin{IEEEbiography}[{\includegraphics[width=1in,height=1.25in,clip,keepaspectratio]{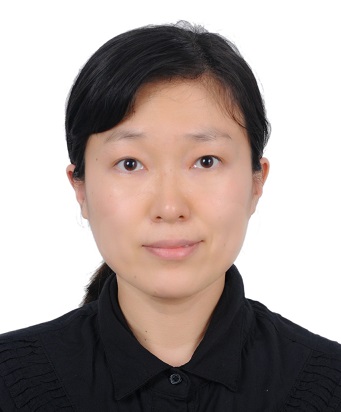}}]{Min Gao}
		received the MS and PhD degrees in computer science from Chongqing University in 2005 and 2010 respectively. She is an associate professor at the School of Big Data \& Software Engineering, Chongqing University. Her research interests include recommendation systems, service computing, and data mining. 
	\end{IEEEbiography}
	
	\begin{IEEEbiography}[{\includegraphics[width=1in,height=1.25in,clip,keepaspectratio]{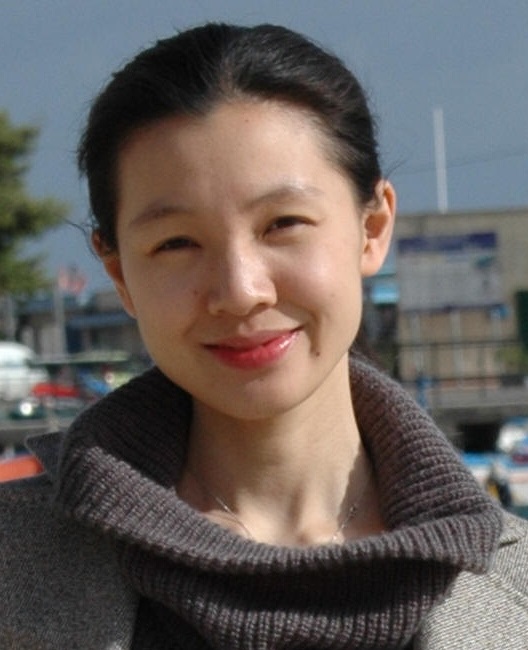}}]{Zi Huang}
		received the BSc degree from the Department of Computer Science, Tsinghua University, China, and the PhD degree in computer science from the School of ITEE, University of Queensland. She is an associate professor (reader) and ARC Future fellow in the School of ITEE, University of Queensland. Her research interests mainly include multimedia indexing and search, social data analysis, and knowledge discovery. 
	\end{IEEEbiography}
	
	\begin{IEEEbiography}[{\includegraphics[width=1in,height=1.25in,clip,keepaspectratio]{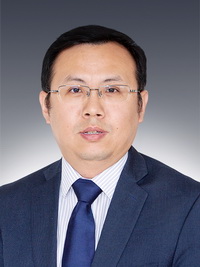}}]{Lizhen Cui}
		is a full professor with Shandong University. He is appointed dean and deputy party secretary for School of Software, co-director of Joint SDU-NTU Centre for Artificial Intelligence Research(C-FAIR), director of the Research Center of Software and Data Engineering, Shandong University. His main interests include big data intelligence theory, data mining, wisdom science, and medical health big data AI applications.
	\end{IEEEbiography}
	%
	%
	%
	
	
	

\end{document}